\documentclass[aps,pre,twocolumn,superscriptaddress,floatfix]{revtex4-2}
\usepackage[colorlinks=true,allcolors=blue, bookmarks=false]{hyperref}
\usepackage{amsfonts,amsmath,amssymb}
\usepackage{graphicx}

\begin{document}

\title{Sensitivity of the spectral form factor to short-range level statistics}

\author{Wouter Buijsman} 

\email{w.buijsman@uva.nl}

\affiliation{Institute for Theoretical Physics Amsterdam and Delta Institute for Theoretical Physics, University of Amsterdam, P.O. Box 94485, 1090 GL Amsterdam, The Netherlands}

\author{Vadim Cheianov}

\affiliation{Instituut-Lorentz and Delta Institute for Theoretical Physics, Universiteit Leiden, P.O. Box 9506, 2300 RA Leiden, The Netherlands}

\author{Vladimir Gritsev}

\affiliation{Institute for Theoretical Physics Amsterdam and Delta Institute for Theoretical Physics, University of Amsterdam, P.O. Box 94485, 1090 GL Amsterdam, The Netherlands}

\affiliation{Russian Quantum Center, Skolkovo, Moscow 143025, Russia}

\date{October 6, 2020}

\begin{abstract}
The spectral form factor is a dynamical probe for level statistics of quantum systems. The early-time behaviour is commonly interpreted as a characterization of two-point correlations at large separation. We argue that this interpretation can be too restrictive by indicating that the self-correlation imposes a constraint on the spectral form factor integrated over time. More generally, we indicate that each expansion coefficient of the two-point correlation function imposes a constraint on the properly weighted time-integrated spectral form factor. We discuss how these constraints can affect the interpretation of the spectral form factor as a probe for ergodicity. We propose a new probe, which eliminates the effect of the constraint imposed by the self-correlation. The use of this probe is demonstrated for a model of randomly incomplete spectra and a Floquet model supporting many-body localization.
\end{abstract}

\maketitle

\section{Introduction}
Level statistics play an unambiguously important role in studies on quantum ergodicity \cite{Haake10, DAlessio16}, thanks to the universal properties as described by random matrix theory \cite{Mehta04, Forrester10}. The applicability of random matrix theory to describe the correlations between energy levels is quantified by the Thouless energy \cite{Edwards72}, which gives the range over which the random matrix theory description for fully ergodic systems holds. For diffusive mesoscopic systems, this range is intimately related to numerous quantities, such as the conductance and the time required for electrons to diffuse over the full sample \cite{Altland94, Beenakker97}. 

The dependence of the Thouless energy on the parameters of the system allows one to study the onset of ergodicity. In the spirit of studies on disordered mesoscopic systems conducted in the 90's \cite{Argaman93, Chalker96, Cohen98}, recent years show a revival of interest for this quantity from several directions \cite{Kos18, Gharibyan18, Chan18, Nosaka18, Friedman19, Sierant20, Sierant20-2, Giudici20, Suntajs20-2}. The correlations within the spectra are typically studied through the spectral form factor \cite{Brezin97, Guhr98}, a time-dependent probe for level statistics on ranges long compared to the mean level spacing.

In this work, we show that the self-correlation imposes a constraint on the spectral form factor integrated over time. More generally, it is shown that each expansion coefficient of the two-point correlation function imposes a constraint on the properly weighted time-integrated spectral form factor. As the lower order expansion coefficients characterize correlations at small separation, the constraints are effectively determined by short-range level statistics. We argue that these constraints can affect the interpretation of the spectral form factor as a probe for long-range level statistics, as well as the usability of the spectral form factor as a tool to study the scaling of the Thouless energy.

We propose a new probe for ergodicity which eliminates the constraint imposed by the self-correlation, thereby providing a more transparent quantification of ergodicity than the spectral form factor. For this probe, quantifying ergodicity does not involve a comparison with the evaluation for fully ergodic systems, giving it the additional benefit that it is applicable even for systems that obey intermediate level statistics. We demonstrate the use of this probe for an ensemble of random incomplete spectra \cite{Bohigas04, Bohigas06} and a Floquet model supporting many-body localization \cite{Zhang16}.

The outline is as follows. Sec. \ref{sec: sff} discusses the spectral form factor and the conventional procedure that is used to study the evolution of the Thouless energy. Sec. \ref{sec: constraints} derives the constraints imposed by short-range level statistics. Sec. \ref{sec: examples} illustrates the potential consequences of these constraints with physically relevant examples from random matrix theory. Sec. \ref{sec: probe} introduces the new probe. Sec. \ref{sec: evaluation} demonstrates the use of the probe for two models of intermediate level statistics. A discussion and outlook is provided in Sec. \ref{sec: discussion}.

\section{Spectral form factor} \label{sec: sff}
We consider an ensemble of spectra $\{ \lambda_n \}_{n=1}^ N$ with $N \gg 1$ levels. The spectra are supposed to be unfolded, meaning that a transformation to unit density is applied. With $\langle \cdot \rangle$ denoting an ensemble average, the density $\rho^{(1)}(x)$ and two-point correlation function $\rho^ {(2)}(\lambda, \lambda')$ are given by
\begin{align}
& \rho^{(1)}(\lambda) = \bigg \langle \sum_n \delta(\lambda - \lambda_n) \bigg \rangle, \\
& \rho^{(2)}(\lambda, \lambda' ) = \bigg \langle \sum_{m \neq n} \delta(\lambda - \lambda_n) \delta(\lambda' - \lambda_m) \bigg \rangle.
\end{align}
Because of the unfolding, one finds $\rho^{(1)}(\lambda) = 1$ over the full range of support. The two-point correlations are assumed to be translationally invariant, meaning that $\rho^ {(2)}(\lambda, \lambda') = \rho^{(2)}(0, \lambda' - \lambda)$. The spectral form factor $K(t)$ \cite{Mehta04} is defined as the Fourier transform of the cluster function $\rho^{(2)}(0,\lambda) - \rho^{(1)}(0) \rho^{(1)}(\lambda)$ accompanied by an offset,
\begin{equation}
K(t) = 1 + \int \bigg( \rho^{(2)}(0,\lambda) - 1 \bigg) e^{i \lambda t} d\lambda.
\label{eq: K}
\end{equation}
Because of the finite range of support for $\lambda$, the time is a discrete variable taking values $2 \pi n / N$ for $n \in \mathbb{Z}$. The translational invariance of the correlations allows one to replace $\exp(i \lambda t)$ by the real-valued $\cos(\lambda t)$. Utilizing the translational invariance of the correlations again, the spectral form factor can be evaluated at relatively low computational costs as
\begin{align}
K(t) 
& = \left \langle \frac{1}{N} \bigg( \sum_{n,m} e^{i (\lambda_n - \lambda_m) t}\bigg) \right \rangle - N \delta(t) \label{eq: K-numerical} \\
& = \left \langle \frac{1}{N} \bigg| \sum_n e^{i \lambda_n t}\bigg|^2 \right \rangle - N \delta(t).
\end{align}
Ensemble averaging is required as the spectral form factor is not a self-averaging quantity \cite{Prange97}. Sec. \ref{sec: evaluation} covers the interpretation of the wavenumber $t$ as a time. 

The large-$\lambda$ behaviour of $\rho^{(2)}(0, \lambda)$ is equivalent to the small-$t$ behaviour of the spectral form factor (restricting the focus to $t \ge 0$), as follows from the Fourier transform of the expansion coefficients,
\begin{equation}
\int \lambda^{-2n} e^{i \lambda t} d \lambda = \frac{\pi (-1)^n}{(2n-1)!} |t|^{2n-1}
\end{equation}
with $n \in \mathbb{N}$ \cite{Lighthill58, Forrester01}. It is suggestive to associate the behaviour of the spectral form factor at time $t$ with the two-point correlator acting over a distance proportional to $1/t$. Then, there is a lowest time from which onwards the spectral form factor matches the evaluation for fully ergodic systems. In the literature, this time known as the Thouless \cite{Kos18, Chan18, Nosaka18, Friedman19, Sierant20, Sierant20-2, Giudici20, Suntajs20-2}, ergodic \cite{Argaman93, Chalker96, Cohen98} or ramp \cite{Gharibyan18} time. It is inversely proportional to the Thouless energy, giving the range over which the spectra obey the correlations as for fully ergodic systems \cite{Edwards72}. We remark that each of Refs. \cite{Kos18, Gharibyan18, Chan18, Nosaka18, Friedman19, Sierant20, Sierant20-2, Giudici20, Suntajs20-2} appeared in recent years.

\section{Constraints imposed by short-range level statistics} \label{sec: constraints}
A key distinction between ergodic and non-ergodic systems is the occurence of level repulsion \cite{Mehta04}. Unfolded spectra obey
\begin{equation}
\rho^{(2)}(\lambda, \lambda) = 
\begin{cases}
0 	& \text{(level repulsion)}, \\
1	& \text{(no level repulsion)}.
\end{cases}
\end{equation}
Integrating the expression for the spectral form factor as given in Eq. \eqref{eq: K} over time shows that the value of the self-correlation $\rho^{(2)}(\lambda, \lambda)$ imposes a constraint on the time-integrated spectral form factor,
\begin{equation}
\int_0^\infty \bigg(1 - K(t) \bigg) dt =\pi \bigg( 1 - \rho^{(2)}(0,0) \bigg). \label{eq: result1}
\end{equation}
For spectra with and without level repulsion, the integral evaluates to respectively $\pi$ and zero. An important consequence appears when determining the Thouless time. Namely, positive (negative) differences between the spectral form factor and the evaluation for fully ergodic systems at earlier times have to be compensated by negative (positive) differences at later times. As such, one could expect the estimated Thouless time to deviate significantly from the value that would have been obtained by using probes that are not sensitive to constraints. 

The constraint imposed by Eq. \eqref{eq: result1} is a specific example from a more general set of constraints when $\rho^{(2)}(0,\lambda)$ can be expanded in powers of $\lambda$. Examples are given in Sec. \ref{sec: examples}. Consider the inverse Fourier transform
\begin{equation}
\rho^{(2)}(0,\lambda) - 1 = \frac{1}{2 \pi} \int \bigg( K(t) - 1 \bigg) e^{-i \lambda t} dt.
\end{equation}
On the left-hand side, we expand $\rho^{(2)}(0,\lambda)$ in powers of $\lambda$ as
\begin{equation}
\rho^{(2)}(0,\lambda) = c_0 + c_1 \lambda^{2} + c_2 \lambda^4 + c_3 \lambda^6 + \dots.
\end{equation}
Next, we take the $2n$-th derivative with respect to $\lambda$ on both sides, and set $\lambda = 0$ to obtain
\begin{equation}
\int_0^\infty \bigg( 1 - K(t) \bigg) t^{2n} dt = \pi (-1)^n (2n)! c_n.
\label{eq: result2}
\end{equation}
This equation establishes a relation between the $2n$-th derivative of $\rho^{(2)}(0, \lambda)$ at $\lambda=0$ and the time-integrated spectral form factor weighted by a factor $t^{2n}$. By a similar argument as above, these constraints can make the estimated Thouless time to deviate from the value that would have been obtained by unconstrained probes.

\section{Illustrations} \label{sec: examples}
Eqs. \eqref{eq: result1} and \eqref{eq: result2} indicate dependencies between the spectral form factor evaluated at earlier and later times. A physically relevant illustration is provided by the bulk statistics of the unitary (Dyson index $\beta = 2$) random matrix ensemble \cite{Dyson62, Mehta04}. In the thermodynamic limit $N \to \infty$, unfolded spectra obey
\begin{align}
\rho^{(2)}(0,\lambda) 
& = 1 - \left( \frac{\sin(\pi x)}{\pi \lambda} \right)^2 \label{eq: rho2-unitary} \\
& = -\sum_{n=0}^ \infty \frac{(-1)^{n} 4^n \pi^ {2n}}{(2n)! (2n^2 + 3n + 1)} \lambda^{2n}. \label{eq: rho2-unitary2}
\end{align}
The level statistics of the unitary random matrix ensemble apply to fully ergodic systems with broken time-reversal symmetry. The constraints given by Eqs. \eqref{eq: result1} and \eqref{eq: result2} impose
\begin{equation}
\int_0^\infty \bigg( 1 - K(t) \bigg) t^{2n} dt = - \frac{4^n \pi^{2n+1}}{2 n^2 + 3n + 1}
\end{equation}
for the $n$-th term of the expansion given in Eq. \eqref{eq: rho2-unitary2}. These constraints are consistent with the evaluation of the spectral form factor \cite{Mehta04}, given by
\begin{equation}
K(t) = 
\begin{cases}
\frac{|t|}{2 \pi} & \text{if } |t| \le 2 \pi, \\
1 & \text{if } |t| > 2 \pi.
\end{cases}
\label{eq: K-unitary}
\end{equation}
Various models for intermediate level statistics, such as the short-range plasma model \cite{Bogomolny99}, are described by Eq. \eqref{eq: rho2-unitary} on short ranges, i.e. up to lower orders. Due to the constraints, the corresponding spectral form factor could match Eq. \eqref{eq: K-unitary} down to early times, despite disagreement of the two-point correlation function with Eq. \eqref{eq: rho2-unitary} at large separations. Again, one might consequently expect the estimated Thouless time to deviate from the value that would have been obtained when using alternative probes that are not sensitive to constraints.

A generalization of the above example is provided by the Calogero-Sutherland model \cite{Sutherland71}. This model is directly related to the random matrix models, see e.g Ref. \cite{Forrester10}. It describes particles on a ring, interacting through a pairwise potential with a magnitude controlled by a parameter $\beta \ge 1$. The positions are indicated by an angle $\theta_n \in [0, 2\pi)$, where $n$ runs over $1,2,\ldots, N$ with $N$ the number of particles. The Hamiltonian $H$ is given by
\begin{equation}
H = - \sum_{n=1}^N \frac{\partial^2}{\partial \theta_n^2} + \sum_{n<m} \frac{\beta (\beta - 2)}{8 \sin^2 \left[(\theta_n - \theta_m) /2 \right]}.
\end{equation}
The amplitude of the ground state matches the joint probability distribution
\begin{equation}
P(\theta_1, \theta_2, \ldots, \theta_N) \sim \prod_{n<m} \left| e^{i \theta_n} - e^{i \theta_m} \right|^\beta
\label{eq: jpdf-beta}
\end{equation}
 of the circular random matrix ensembles, where $\beta$ denotes the Dyson index \cite{Mehta04}. The Calogero-Sutherland model thus generalizes the Dyson index from $\beta \in \{1,2,4\}$ to the continues range $\beta \ge 1$. Eq. \eqref{eq: result2} imposes a constraint on the time-integrated spectral form factor when $\beta$ is even (i.e. $\beta / 2 \in \mathbb{N}$). In the thermodynamic limit $N \to \infty$, these coefficients have been obtained analytically in terms of generalized factorials \cite{Forrester92}. Noting that the Hamiltonian of the Calogero-Sutherland model is integrable, we conjecture that these coefficients are related to the conserved charges. For $\beta = 4$ (symplectic random matrix ensemble), the coefficients can alternatively be obtained from the evaluation of the spectral form factor \cite{Mehta04}, given by
\begin{equation}
K(t) = 
\begin{cases}
\frac{|t|}{4\pi} - \frac{|t|}{8 \pi} \ln \bigg| 1 - \frac{|t|}{2 \pi} \bigg| & \text{if } |t| \le 4 \pi, \\
1 & \text{if } |t| > 4 \pi.
\end{cases}
\label{eq: K-symplectic}
\end{equation}

\section{Proposal for a new probe} \label{sec: probe}
The evolution of the Thouless time as a function of the parameters of the system allows one to quantify the onset of ergodicity through the spectral form factor. Because of the constraints outlined in Sec. \ref{sec: constraints}, the interpretation can however be fairly non-straightforward. Eq. \eqref{eq: result1} suggests an alternative probe, defined as
\begin{equation}
\rho^{(2)}(0,0|t) = 1 - \frac{1}{\pi} \int_0^t \bigg(1 - K(t') \bigg) dt'.
\label{eq: probe}
\end{equation}
This probe gives the self-correlation as captured by the spectral form factor evaluated at times less than $t$. Poissonian level statistics are characterized by $K(t) = 1$ as the levels are uncorrelated. With increasing $t$, the value of $\rho^{(2)}(0,0|t)$ thus tends to zero or one for respectively ergodic and non-ergodic systems. It approximates the self-correlation in a controlled way, thereby serving as a diagnostic when evaluated at a fixed, late time.

The diagnostic proposed here is advantageous compared to the Thouless time in at least two respects. First, the constraint on the time-integrated spectral form factor imposed by the self-correlation through Eq. \eqref{eq: result1} is eliminated, making it arguably more transparent. Second, the quantification of ergodicity is not based on a comparison with fully ergodic systems, thereby allowing one to study systems exhibiting intermediate level statistics. 

Semiclassically, the Heisenberg time $t_\text{H} = 2 \pi$ is the largest physically relevant time (see e.g. Refs. \cite{Muller04, Heusler07}). In this setting, it can be natural to quantify the onset of ergodicity by $\rho^{(2)}(0,0|t_\text{H})$. Using the evaluations of the spectral form factors for the bulk statistics of the orthogonal ($\beta = 1$), unitary ($\beta = 2$) and symplectic ($\beta = 4$) random matrix ensembles as given in respectively Eqs. \eqref{eq: K-orthogonal}, \eqref{eq: K-unitary}, and \eqref{eq: K-symplectic}, one finds
\begin{equation}
\rho^{(2)}(0,0|t_\text{H})
 = 
\begin{cases}
1 - \frac{3}{4}\ln(3) & \text{if } \beta = 1, \\
0 & \text{if } \beta = 2, \\
0 & \text{if } \beta = 4, \\
\end{cases}
\label{eq: rho0tH}
\end{equation}
with $1 - \frac{3}{4}\ln(3) \approx 0.176$. We note that $t_\text{H} = 4 \pi$ for the symplectic ensemble as the spectra contain $2N$ elements \cite{Mehta04}. A method to numerically evaluate the integral in Eq. \eqref{eq: probe} up to arbitrarily large times at low computational costs is mentioned in Sec. \ref{sec: evaluation}.

\section{Examples} \label{sec: evaluation}

\subsection{Randomly incomplete spectra}
Randomly incomplete spectra provide an interpolation between Poissonian and Wigner-Dyson level statistics, for which the spectral form factor can be obtained analytically \cite{Bohigas04}. Following Ref. \cite{Bohigas06}, we consider the bulk statistics of the orthogonal ($\beta = 1$) random matrix ensemble with a randomly selected fraction $1-f$ of the levels omitted. This ensemble was introduced originally to study the effect of missing levels in experimental contexts. After rescaling the levels to unit mean level spacing (unfolding), the spectral form factor $K(t)$ is given by
\begin{equation}
K(t) = 1 - f + f K'(f t),
\label{eq: K-ris}
\end{equation}
where $K'(t)$ denotes the spectral form factor for the bulk statistics of the orthogonal random matrix ensemble \cite{Mehta04},
\begin{equation}
K'(t) = 
\begin{cases}
\frac{|t|}{\pi} - \frac{|t|}{2 \pi} \ln \bigg( 1 + \frac{|t|}{\pi} \bigg) & \text{if } |t| \le 2 \pi, \\
2 - \frac{|t|}{2 \pi} \ln \bigg( \frac{|t| / \pi + 1}{|t| / \pi - 1} \bigg) & \text{if } |t| > 2 \pi.
\end{cases} \label{eq: K-orthogonal}
\end{equation}
The ensemble interpolates between Poissonian ($f=0$) and Wigner-Dyson ($f=1$) level statistics. Level repulsion can be observed for $f>0$. For the intermediate value $f=1/2$, the statistics are close to those of the semi-Poisson model \cite{Bogomolny02}. Evaluating $\rho^{(2)}(0,0|t)$ at the Heisenberg time $t = 2 \pi$ yields
\begin{equation}
\begin{split}
& \rho^{(2)}(0,0|t_H) = 1 + \frac{5}{2}(f^2-f) + \left(\frac{1}{4} - f^2 \right) \ln(1+2f).
\end{split}
\end{equation}
Consistent with Eq. \eqref{eq: rho0tH}, this evaluates to respectively unity and $1 - \frac{3}{4} \ln(3)$ for $f=0$ and $f=1$.

Eq. \eqref{eq: K-ris} gives the spectral form factor for a weighted sum of the complete spectra (factor $f$, combined with a scaling of the density) and uncorrelated levels (factor $1-f$). Superimposed spectra appear more frequently as models for intermediate level statistics \cite{Drozdz91}. Potentially, alternative interpolations can be obtained from superimposed spectra.

\subsection{Many-body localization}
Quantum systems with time-periodic Hamiltonians are known as Floquet systems \cite{Shirley65}. The Floquet operator $U_F$ is given by the time-evolution operator of the Hamiltonian $H(t)$ acting over a single period $T$,
\begin{equation}
U_F = \exp \bigg(- i \hbar \int_0^T H(t) dt \bigg).
\end{equation}
Since $U_F$ is unitary, the eigenvalues can be parametrized as $e^{i \theta}$ with $0 \le \theta < 2 \pi$. The set $\{ \theta_n \}$ gives the quasi-energy spectrum. The set of quasi-energy levels of the $n$-th power of $U_F$, which is the time-evolution operator for $n$ cycles, is given by $\{\theta_i n \} = \{x_i t \}$. As $t$ only enters in the expression for the spectral form factor as $\theta t$, it has the interpretation of a (discrete) time.

We consider the Floquet model introduced in Ref. \cite{Zhang16}. It describes a spin-$1/2$ chain subject to disorder and an external field switching back-and-forth between the two directions. The Floquet operator is given by
\begin{align}
& U_F = \exp(- i \tau H_x) \exp (-i \tau H_z), \\
& H_x = \sum_{n=1}^{L} g \Gamma \sigma_n^x, \\
& H_z = \sum_{n=1}^{L-1} \sigma_n^z \sigma_{n+1}^z + \sum_{n=1}^{L} \left(h+g\sqrt{1-\Gamma^2} G_n \right) \sigma_n^z.
\end{align}
The $\sigma_n^{x,z}$ represent Pauli matrices acting on site $n$. Periodic boundary conditions $\sigma^{x,z}_{L+1} = \sigma^{x,z}_1$ are imposed. The $G_n$ represent disorder sampled independently from a Gaussian distribution with mean zero and unit variance. The free parameters are taken as $g=0.9045$, $h = 0.809$, and $\tau = 0.8$. The spectral density is uniform. The results below are obtained from statistics over at least $1000$ disorder realizations. The model exhibits many-body localization \cite{Ponte15, Ponte15-1, Lazarides15, Abanin16, Abanin19}. At large $L$, it is indicated to be in a localized phase for $\Gamma \lesssim 0.3$ (see also Ref. \cite{Lezama19}). 

\begin{figure}
\centerline{\includegraphics[scale=0.8]{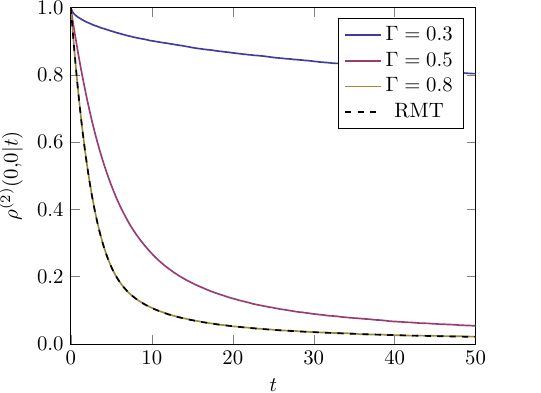}}
\caption{Numerically obtained $\rho^{(2)}(0,0|t)$ as a function of $t$ at system size $L=12$ for $\Gamma = 0.3$ (top curve), $\Gamma = 0.5$ (middle curve), and $\Gamma = 0.8$ (lower curve). The dashed line displays the random matrix (RMT) theory prediction for fully ergodic systems.}
\label{fig: probe-t}
\end{figure}

\begin{figure}
\centerline{\includegraphics[scale=0.8]{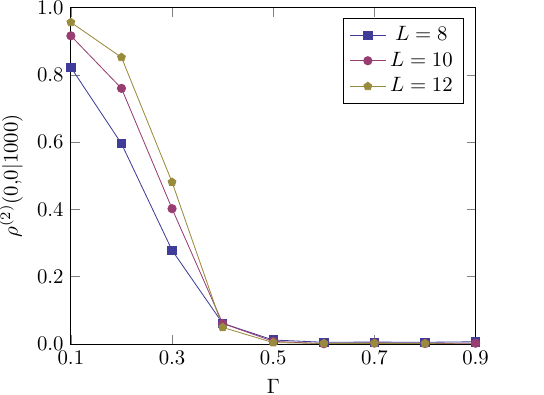}}
\caption{Numerically obtained $\rho^{(2)}(0,0| 1000)$ as a function of $\Gamma$ for $L=8$ (squares), $L=10$ (circles), and $L=12$ (pentagons). For Poissonian and Wigner-Dyson level statistics, one expects respectively $\rho^{(2)}(0,0| 1000) = 1$ and $\rho^{(2)}(0,0| 1000) \approx 0.001$.}
\label{fig: probe-Gamma}
\end{figure}

Fig. \ref{fig: probe-t} shows the evolution of $\rho^{(2)}(0,0|t)$ as a function of $t$. In the ergodic phase ($\Gamma = 0.8$), one observes strong agreement with the random matrix theory evaluation, obtained from the spectral form factor given in Eq. \eqref{eq: K-orthogonal}. When approaching the localized phase, the curve tends towards $K(t) = 1$ for Poissonian level statistics. Fig. \ref{fig: probe-Gamma} shows the evolution of $\rho^{(2)}(0,0|1000)$ as a function of $\Gamma$. As in Refs. \cite{Zhang16, Lezama19}, the results are consistent with a transition between an ergodic and a non-ergodic phase at $\Gamma \approx 0.3$ at large system sizes. The integral in the evaluation of $\rho^{(2)}(0,0|t)$ can be evaluated at relatively low computational costs by using Eq. \eqref{eq: K-numerical} and involving
\begin{equation}
\sum_{n=0}^N \cos(n x) = \frac{1}{\sin(x/2)} \sin \left( \frac{x}{2}(N+1) \right) \cos \left( \frac{x}{2} N \right).
\end{equation}

\section{Discussion and conclusions} \label{sec: discussion}
In summary, we revisited the interpretation of the spectral form factor as a probe for ergodicity. We have shown that short-range level statistics imposes constraints on the spectral form factor integrated over time, which could affect its interpretation as a probe for long-range level statistics, as well as the usability of the spectral form factor as a tool to study the scaling of the Thouless energy. We have proposed a new probe, and argued that it is more transparent. We demonstrated the use of this probe for two models of intermediate level statistics.

The Thouless energy can alternatively be determined from the number variance \cite{Weidemuller09}. Interestingly, this yields results conflicting with the analysis of the spectral form factor for several classes of systems \cite{Gharibyan18}. Possibly, these discrepancies can be explained using the results of this paper. Next, our probe could be relevant in the recently emerging debate on the stability of many-body localization \cite{Kiefer-Emmanouilidis20, Suntajs20, Suntajs20-2, Luitz20}, in which the spectral form factor plays a prominent role. We remark that the spectral form factor appears in the survival probability for fully ergodic systems \cite{Torres-Herrera18, Schiulaz19}. Finally, we note that the newly proposed probe broadens the range of usability to systems displaying intermediate level statistics.

\begin{acknowledgments}
This work is part of the Delta-ITP consortium, a program of the Netherlands Organization for Scientific Research (NWO) that is funded by the Dutch Ministry of Education, Culture and Science (OCW).
\end{acknowledgments}

\bibliography{formfactor-v2}

\end{document}